\documentclass[aps,prb,reprint,showpacs,floatfix,citeautoscript,superscriptaddress]{revtex4-1}
\usepackage[latin1]{inputenc}
\usepackage{color}
\usepackage{dcolumn}
\usepackage[pdftex]{graphicx}
\usepackage{fixmath}
\usepackage{xspace}
\usepackage[colorlinks=true,linkcolor=blue,%
citecolor=blue,urlcolor=blue,pdfborder={0 0 0}]{hyperref}

\usepackage{textcomp}
\usepackage{amssymb}
\usepackage{amsmath}
\usepackage{amsthm}
\usepackage{bbm}
\usepackage{bm}
\usepackage{multirow}
\usepackage{psfrag}
\usepackage{mathrsfs}
\usepackage{footmisc}
\usepackage[caption=false]{subfig}
\usepackage{siunitx}
\usepackage{tabularx}

\newcommand{\etal}{\emph{et al.}\xspace}

\newcolumntype{d}{D{.}{.}{0}}
\newcolumntype{x}{D{,}{\times}{-1}}
\newcolumntype{Y}{>{\centering\arraybackslash}X}

\newcommand{\udem}{D\'{e}partement de Physique and Regroupement Qu\'{e}b\'{e}cois sur les Mat\'{e}riaux de Pointe (RQMP), Universit\'{e} de Montr\'{e}al, C.P. 6128, Succursale Centre-Ville, Montr\'{e}al, Qu\'{e}bec, Canada H3C~3J7} 
\newcommand{\warwick}{Department of Physics and Centre for Scientific Computing, University of Warwick, Gibbet Hill Road, Coventry CV4~7AL, United Kingdom}
 \newcommand{\ORNL}{Materials Science and Technology Division, Oak Ridge National Laboratory, Oak Ridge, Tennessee 37831-6138, USA}
 \newcommand{\Carleton}{Department of Mechanical and Aerospace Engineering, Carleton University, 1125 Colonel By Drive, Ottawa, Ontario, Canada K1S~5B6}

\begin{document}

\title{Understanding long-time vacancy aggregation in iron: a kinetic Activation-Relaxation Technique study}

\author{{Peter} \surname{Brommer}}
   \email{p.brommer@warwick.ac.uk}
  \altaffiliation[Current address: ]{Centre for Predictive Modelling, School of Engineering, Library Road, Coventry CV4 7AL, United Kingdom}
  \affiliation{\udem}\affiliation{\warwick}
\author{{Laurent Karim} \surname{B\'eland}}
\altaffiliation[Current address: ]{\ORNL}
  \affiliation{\udem}
\author{{Jean-Fran\c{c}ois} \surname{Joly}}
\altaffiliation[Current address: ]{\Carleton}
  \affiliation{\udem}
\author{{Normand} \surname{Mousseau}}
   \email{normand.mousseau@umontreal.ca}
   \affiliation{\udem}

\date{\today}


\begin{abstract}

Vacancy diffusion and clustering processes in body-centered cubic (bcc) Fe are studied using the kinetic Activation-Relaxation Technique (k-ART), an off-lattice kinetic Monte Carlo (KMC) method with on-the-fly catalog building capabilities. 
For mono- and di-vacancies, k-ART recovers previously published results while clustering in a 50-vacancy simulation box agrees with experimental estimates. 
Applying k-ART to the study of clustering pathways for systems containing from 1 to 6 vacancies, we find a rich set of diffusion mechanisms. 
In particular we show that the path followed to reach a hexavacancy cluster influences greatly the associated mean-square displacement. 
Aggregation in a 50-vacancy box also shows a notable dispersion in relaxation time associated with effective barriers varying from 0.84 to 1.1 eV depending on the exact pathway selected.  
We isolate the effects of long-range elastic interactions between defects by comparing to simulations where those effects are deliberately suppressed.
This allows us to demonstrate that in bcc Fe, suppressing long-range interactions mainly influences kinetics in the first 0.3 ms, slowing down quick energy release cascades seen more frequently in full simulations, whereas long-term behavior and final state are not significantly affected.
\end{abstract}

\maketitle

\section{Introduction}
\label{sec:introduction}

The study of irradiation damage is central to understanding materials kinetics.
Ion bombardment leaves behind defect cascades of self-interstitial atoms (SIA) and vacancies that can migrate and affect the mechanical properties of the materials, forming cracks or swelling, for example.\cite{DiazdelaRubia:2000:871,Fu:2005:68} 
Because of the microscopic nature of these processes, numerical simulations are needed to provide crucial understanding about atomistic details that is difficult to obtain from experiments.
While simulations have contributed significant information on this topic, they have also been limited due to the extended time scale over which many of these processes take place; a time scale that is out of reach from standard molecular dynamical simulations.
Kinetic Monte Carlo (KMC)\cite{Bortz:1975:10} provides a solution to reaching long time dynamics, however the standard implementation requires an upfront knowledge of the relevant barriers and cannot take into account crucial elastic deformations.\cite{Fu:2005:68} 
While results from such simulations are enlightening, their quantitative validity is limited since the full details of local atomic configurations can affect greatly diffusion kinetics.\cite{Djurabekova:2010:2585} 
Over the last few years, numerous improvements have been proposed to standard KMC simulations, in order to overcome these limitations\cite{El-Mellouhi:2008:153202,Kushima:2009:224504,Xu:2012:375402} (see Ref.~\onlinecite{Mousseau:2012:925278} for additional references), leading to a new interest in the fundamental mechanisms associated with defect diffusion in nuclear materials.

Radiation-damage recovery in Fe is generally described in five stages.\cite{Takaki:1983:87,Fu:2005:68} 
At low temperature, the first two stages are associated with interstitial--vacancy (I--V) pair recombination; the third one, below 200 K, with di-interstitial diffusion, the fourth one, near room temperature, with vacancy diffusion and the final one, at 520--550 K, with defect-cluster dissociation.
In spite of this apparent clean energy differentiation between the various stages, a number of recent results suggest that damage recovery is a more complex process, where the various defects play an intertwined role.\cite{Marinica:2011:94119,Marinica:2012:25501} 

Using the kinetic Activation-Relaxation-Technique (k-ART), an off-lattice kinetic Monte-Carlo method with on-the-fly catalog building capabilities, we revisit this issue in order to clarify these mechanisms. 
K-ART allows us to reach experimental time scales, many orders of magnitude longer than reachable by molecular dynamics while incorporating exact elastic effects and identifying the atomistic details of diffusion mechanisms at every step. 
In this paper, we focus on vacancies, which dominate diffusion above room temperature. 
This is also the temperature range where KMC techniques, which move the system across a potential energy surface determined at \SI{0}{\kelvin}, are the method of choice.
For simulations at higher temperatures, when the time scales accessible to KMC decrease due to increasing reaction rates, alternative methods like Molecular Dynamics (MD) may be better suited to the task.
At the temperatures under study here, following diffusion mechanisms and pathways allows us to show that we can recover previous simulation results and provide a number of new insights regarding both the diffusion process of small vacancy clusters and the aggregation phase.

\section{Methods}
\label{sec:methods}

Simulations presented here are based on kinetic ART,\cite{El-Mellouhi:2008:153202,Beland:2011:046704} a powerful and versatile algorithm with on-the-fly event catalog building capacity and exact treatment of elasticity.
K-ART uses the activation-relaxation technique (ART nouveau) for event searching,\cite{Barkema:1996:4358,Malek:2000:7723,Marinica:2011:94119,Machado-Charry:2011:34102} an efficient saddle-point searching method that was shown to be particularly effective for finding events in iron\cite{Marinica:2011:94119}.
Combining ART nouveau with a topological analysis package\cite{McKay:1981:45}, k-ART is not limited to crystalline environments and it can classify configurations and events with any level of disorder.
Over the last few years, k-ART was applied with success to complex systems such as ion-bombarded\cite{Beland:2013:105502} and amorphous silicon.\cite{Joly:2013:144204} 
A description of the method can be found in Ref.~\onlinecite{Beland:2011:046704}.
We present here only the basic steps associated with a k-ART run.

\begin{enumerate}
  \item Starting from a local minimum, we first determine the topology associated with the local environment of each atom in the cell.\label{step1}
  \item We then verify in the event catalog whether each topology currently present has been already observed and has been sufficiently searched for events.
  \item If not, ART nouveau searches are launched on the topologies to complete the catalog; if yes, we move directly to the next step.
  \item All events associated with the topologies characterizing the configuration are placed in the list of available events for this local minimum.
  \item Events that account for at least 99.99\% of the total rate are fully reconstructed and their saddle point fully relaxed to include all elastic deformations associated with the specific configuration.
  \item The total rate is computed again, the clock is moved forward according to a Poisson distribution, an event is selected at random with the appropriate weight.
  \item The event is applied and the configuration is relaxed into a new minimum. 
    We can go back to step \ref{step1}. 
\end{enumerate}

We use a 2000-atom body-centered cubic (bcc) Fe crystal (10$\times$10$\times$10 cubic unit cells) with periodic boundary conditions and the Ackland-Mendelev effective interaction potential A04\cite{Ackland:2004:2629}.
A number of simulations are also performed with the potential M07 by Malerba, Marinica, \emph{et al.}\cite{Malerba:2010:19} in order to assess the influence of the effective potential used.
We remove between one and six atoms to focus on the vacancy diffusion associated with the fourth stage in radiation-damaged recovery.
We also remove 50 atoms to compare vacancy-aggregation results with other simulations\cite{Fan:2011:125501,Brommer:2012:219601, Fan:2012:219602,Xu:2013:66} and positron-annihilation experiments.\cite{Eldrup:2003:346}

All simulations are performed with a constant prefactor of $5\cdot 10^{12}$ s$^{-1}$, which was shown to be a good approximation\cite{Yildirim:2007:165421,Stroscio:1994:8522,Papanicolaou:2009:1366,Fan:2011:125501}.
We use a radius of 5.6~\AA\ for the topological classification, with an initial 50 searches for a new topology.
Additional searches are launched based on logarithmic increase in the number of times this topology is observed, i.e.~a topology appearing $n$ times is searched $50 (1+\log_{10}n)$ times in total.
We ignore events associated with perfectly crystalline environments as barriers are too high to be selected on the simulation time scale.
For the initial configuration of the 50-vacancy system, we identify between 616 and 766 different generic events. 
By the time the simulation reaches 1 ms, we will have generated tens of thousands of events (between 38056 and 82924 events for the four runs).
Comparing with other off-lattice KMC simulations performed on the same system but with a different algorithm (SEAKMC\cite{Xu:2012:375402}), we observe that both approaches lead to similar time scales for vacancy aggregation, suggesting that both methods manage to generate physically relevant pathways, as discussed in Sec.~\ref{sec:50-vacancy-aggr}.\cite{Xu:2012:102,Xu:2013:66}

The first simulations are run at 573 K to allow an essential comparison with earlier lattice KMC results and analytical estimations.\cite{Borodin:2007:161}
Aggregation of large clusters is performed near room temperature, in the vacancy-dominated temperature regime.

\subsection{Handling small barriers}
\label{sec:handl-small-barr}

A major obstacle to overcome in achieving efficient KMC simulations is the problem of small barriers, which plagues many KMC simulation techniques. 
If a system has a basin, i.e.~a group of states separated by energy barriers that are significantly lower than those connecting the basin to other states, then standard KMC will show frequent jumps within the basin (so-called flickers).
Such flickers do not progress the simulation and provide little insight beyond the first tour through those states.
To avoid these fruitless KMC steps, k-ART uses the basin-autoconstructing Mean Rate Method (bac-MRM)\cite{Puchala:2010:134104,Beland:2011:046704}.  
With bac-MRM, final states of events with both forward and backward barriers below a user-specified basin threshold are dynamically added to the current basin.
K-ART then averages over all possible transitions within a basin to pick the next step.
This ensures that each k-ART step advances the simulation by either expanding a basin (until its complete exploration) or by leaving the basin behind.
MRM-like acceleration methods have successfully been used recently in other KMC codes\cite{Fichthorn:2013:164104}.

Because the low-barrier treatment is statistically exact, the choice of the low-barrier threshold does not affect the system's kinetics.
Since particular trajectories are lost within the basin, however, it is preferable to use the lowest threshold possible for specific systems in order to preserve atomistic details about these.
Typically, the barrier threshold needs to be increased over the course of a vacancy clustering simulation, as flickers on growing energy scales become significant obstacles.
For example, in the 50-vacancy clustering simulations, we start with initial threshold of 0.152 eV, increasing it to 0.75 eV as the system evolves (see also Sec.~\ref{sec:50-vacancy-aggr} for details).

\subsection{Generic embedding: eliminating long range effects}
\label{sec:gen_emb}

To study the significance of long-range interactions between defects, we created a modified version of k-ART that intentionally suppresses all structural information in barrier determination beyond the local environment used for topological identification.
In a normal k-ART run, these elastic interactions are accounted for during event reconstruction (step 5 in the sequence described in Sec.~\ref{sec:methods} above), when the most important events for a certain topology are refined for each atom that shares that topology.
These atoms have the same local environment up to the topology cutoff radius, but may differ beyond that.
A crude approach to ignore these differences would be to just skip the refinement step, but this would introduce a bias, as every single catalog event was found in a certain configuration of all atoms.

To counter this problem, we introduce \emph{generic embedding} (GE): We launch the ART nouveau searches (step 3 in the k-ART method) in a modified structure, where the atoms within the sphere used for topological classification (5.6~\AA\ radius) are embedded in a defect-free crystal of bcc iron.
The events found in this way are free from any hidden bias caused by atomic configurations outside the environment classification.
These events are then used without further refinement (which would defy the purpose of GE).
The final configuration of a GE k-ART event is relaxed into the new minimum as in the standard k-ART recipe.

GE simulations are computationally less demanding than full k-ART simulations.
The fact that no event reconstruction is necessary far outweighs the overhead of creating the embedding environment, which is negligible compared to performing an ART nouveau search in that structure.
During the early simulation phases of a system far from equilibrium, when there are many new local environments introduced after almost every step, both methods use a comparable amout of CPU time per KMC step.
But as soon as the system  at least temporarily reaches a stationary state with no new environments to search, GE can perform a KMC step in a few CPU-minutes, whereas the barrier refinement limits full k-ART to about 1 CPU-hour per KMC step for the simulations described in Sec.~\ref{sec:50-vacancy-aggr} below. 
Overall, those GE simulations were a factor 4--5 faster than their full k-ART counterparts for a comparable number of KMC steps.

\subsection{Force field}
\label{sec:forcefield}

As k-ART simulations, like other improved or adaptive KMC simulations, require a large number of force evaluations to build a catalog (on the order of 300 to 600 per successful saddle point search\cite{Mousseau:2012:925278}), they are only feasible with a classical effective potential or force field, where the interactions between electrons and nuclei is reduced to an effective interaction between atoms.
Such simplification comes at a cost: While ab-initio calculations using density functional theory (DFT) methods do not in principle require any additional input, the force field parameters need to be determined beforehand.
The parameters are usually obtained by optimizing certain physical properties (like activation barriers) within the limit of the selected model.
However, this implies that there is a loss of generality; properties \emph{not} included in the optimization process may not be equally well reproduced.

Iron is typically described by embedded atom method (EAM) potentials\cite{Daw:1984:6443}, \begin{equation}
  \label{eq:EAM}
  V=\sum_iU\left[\sum_j\rho(r_{ij})\right]+\sum_{ij}\phi(r_{ij}),
\end{equation}
where $U[n]$ is the embedding energy of an atom in a local density $n=\sum_j\rho(r_{ij})$, defined as the sum of the contributions from neighbors at distance $r_{ij}$ via a transfer function $\rho(r_{ij})$. $\phi(r_{ij})$ is a pair potential function. The properties of the material in simulation is determined by the choices for the three functions $U(n)$, $\rho(r_{ij})$ and $\phi(r_{ij})$.

One of the first Fe potentials created specifically for the simulation of defects was by Ackland, Mendelev, \etal (referenced as A04)\cite{Ackland:2004:2629}, based on an earlier version by the same authors\cite{Mendelev:2003:3977}.
While these potentials did not include the vacancy migration barrier energy in the fit, they can reproduce the activation energy for self-diffusion, which is the sum of vacancy formation and migration energies, to within 10 \% of the experimental value.\cite{Mendelev:2003:3977} For this potential, the energy profile for vacancy migration is double-humped, i.e.~there is a minimum at the midpoint of the trajectory, and the maximum is slightly displaced towards either end.
This is an artifact of the potential; DFT calculations do not show this intermediate minimum\cite{Malerba:2010:19}.
This is why we also use here a potential developed by Malerba, Marinica, \emph{et al.}~(referenced as M07)\cite{Malerba:2010:19}, which uses vacancy formation and migration energies directly in the fitting procedure, removing the double-hump problem.
M07 also yields a slightly higher value for the migration barrier (0.68 eV compared to 0.63 eV with A04), in agreement with DFT calculations.
This difference directly influences the time scales for vacancy kinetics and will be addressed in Sec.~\ref{sec:small-vacancy-clust}.

\section{Results}
\label{sec:results}

\subsection{Small vacancy clusters}
\label{sec:small-vacancy-clust}

To establish a reference point, we first examine the monovacancy using the A04 potential.
At low temperature, in simulations over 5000 KMC steps corresponding to a time scale on the order of 1 s at 300 K, the vacancy first nearest neighbor (1NN) jump (0.64 eV) takes place directly or via a split vacancy position (0.12 eV below saddle point barrier), with 100 \% probability.
The barrier heights agree with literature values\cite{Malerba:2010:19}.
As the temperature rises above 700 K, the next highest barrier, corresponding to the third nearest neighbor (3NN) split vacancy (1.07 eV), starts occurring (0.1 \% at 700 K and 1.3 \% at 1200 K); given the energy difference between the lowest barrier and the next one, the diffusion remains controlled by the 0.64 eV barrier even at high temperature.

In a recent comparative study of effective potentials for iron, Malerba \etal\cite{Malerba:2010:19} report an artificial local minimum in the vacancy 1NN diffusion barrier that is not found in first-principles simulations (cf.\ Sec.~\ref{sec:forcefield}).
This local minimum corresponds to the split vacancy position found in our simulations with the A04 potential. 
The authors suggest a modified potential referenced as M07 that does not show this artifact, however its 1NN vacancy diffusion barrier is slightly higher at 0.68 eV.
At 573 K this difference in barrier height corresponds to a rate that is lower by a factor of 2.7.
Additionally, the A04 k-ART simulation finds twice the number of events with the 1NN barrier for a single vacancy, as it identifies the direct process and the artificial one via an intermediate split vacancy as physically distinct events.
This accounts for a total rate difference of factor 5.4 between M07 and A04.
As the use of the A04 potential is prevalent in point defect simulations in iron,\cite{Gordon:2005:214104,Djurabekova:2010:2585,Fan:2010:104102,Fan:2011:125501,Brommer:2012:219601,Fan:2012:219602,Bjorkas:2012:24105,Xu:2012:375402,Xu:2013:66} we will also use it for reasons of comparability, but will compare with results obtained with M07 in certain cases.

\begin{figure}[hb]
  \includegraphics[width=\columnwidth]{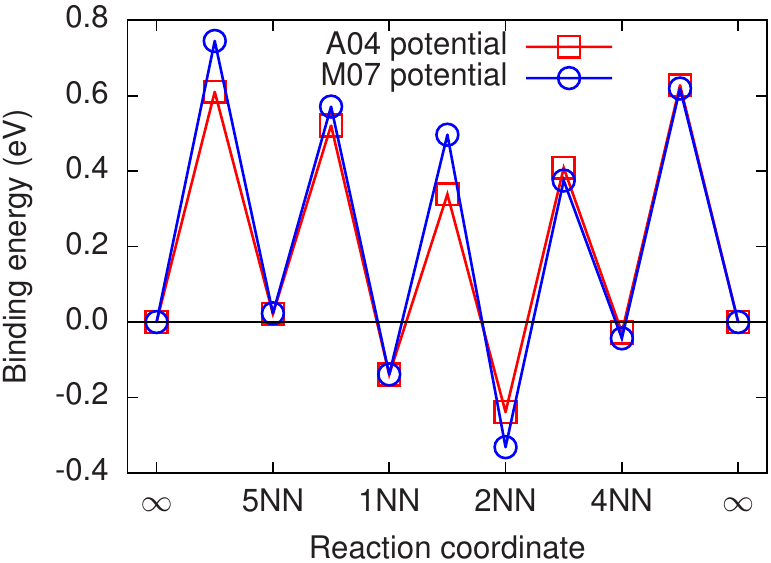}
  \caption{(Color online) Energy landscape of the divacancy states and saddle points according to the A04 and M07 force fields. 
    The energy zero corresponds to two isolated vacancies ($\infty$).\label{fig:2vac_traj} }
  
\end{figure}

For a divacancy, we find that both A04 and M07 reproduce three distinct bound states at 4NN, 1NN, and 2NN position (in order of increasing binding energy).
The energies of divacancy states and connecting saddle points are shown in Fig.~\ref{fig:2vac_traj}.
DFT calculations predict an additional bound state at the 5NN position, but this is usually not reproduced by an effective interaction potential.\cite{Djurabekova:2010:2585} Our results for A04 agree with a previous comparison between various force fields and first-principles results in Ref.~\onlinecite{Djurabekova:2010:2585} (where the A04 potential is called AM); the M07 was not part of this study.
The most significant difference between M07 and A04 is that the 2NN state is more strongly bound by 0.1 eV for M07 with barriers to the 1NN (4NN) state higher by 0.25 (0.06) eV, respectively.
This implies that the M07 favors a divacancy propagation via an intermediate 4NN state rather than a 1NN state, while still being more strongly bound overall.
At a simulation temperature of 573 K, the divacancy dissociates within a few tens of KMC steps.

\begin{table}
  \caption{Lifetimes and diffusion constants of vacancy clusters at 573 K for the A04 \cite{Ackland:2004:2629} and M07 \cite{Malerba:2010:19} potentials. 
    The A04 divacancy dissociates before a diffusion coefficient can be measured.}
  \label{tab:difflife}
  \begin{ruledtabular}
    \begin{tabular}{ldddd}
      Cluster & \multicolumn{2}{c}{Thermal lifetime (s)} & \multicolumn{2}{c}{Diffusion coefficient}\\
                 &                           &                                  &\multicolumn{2}{c}{(nm$^2$ s$^{-1}$)}\\
                 &\multicolumn{1}{c}{A04}&\multicolumn{1}{c}{M07}&\multicolumn{1}{c}{A04}&\multicolumn{1}{c}{M07}\\\colrule
      v         &\multicolumn{2}{c}{N/A}                       & 1\times 10^7 & 2\times 10^6\\
      v$_2$  & 3\times 10^{-8} & 5\times 10^{-7} &\multicolumn{1}{c}{--}&  2\times10^5\\
      v$_3$  & 6\times 10^{-8} &2\times 10^{-6} &  2\times 10^7 &4\times 10^6\\
      v$_4$  & 1\times 10^{-6} &1\times 10^{-4} &  2\times 10^6 &4\times 10^4 \\
      v$_5$  & 5\times 10^{-6} &2\times 10^{-3} &  7\times 10^4 &1\times 10^3\\
      v$_6$  & 4\times 10^{-5} &4\times 10^{-2} &  5\times 10^3 &3\times 10^1\\
    \end{tabular}
  \end{ruledtabular}
\end{table}

Next, we study lifetime and diffusivity of small vacancy clusters (1--6 vacancies).
These simulations are performed in a 2000-atom box at 573 K. 
The vacancies are initially placed in the lowest-energy configuration for a vacancy cluster of that size.
A cluster is deemed decayed, if at least one of its vacancies moves further away than the 4NN site from each of the remaining cluster vacancies.
If a cluster re-forms during a simulation, the lifetime is sampled another time.
The diffusion coefficient quantifies the diffusion of the cluster as a whole during its lifetime.
The results can be found in Table \ref{tab:difflife}. 
For mono- and trivacancies, the diffusion coefficients for A04 are higher than those of M07 by a factor of 5, which agrees with the rate difference between the two potentials. 
In larger clusters, the results from the two potentials diverge: M07 yields significantly longer lifetimes and lower diffusion coefficients than A04.
This may be due to the artificial minimum near the saddle point between two nearest neighbor vacancy positions, and thus be an artifact of the potential.
In agreement with Ref.~\onlinecite{Borodin:2007:161}, we find that the 3-vacancy has an exceptionally high diffusion coefficient.
This is due to an extremely effective migration mechanism with a low activation barrier (cf.\ Ref.~\onlinecite{Fu:2005:68}).

\begin{figure}[hb]
  \subfloat[Mobile configuration v$_3^m$]{%
    \includegraphics[width=4.3cm]{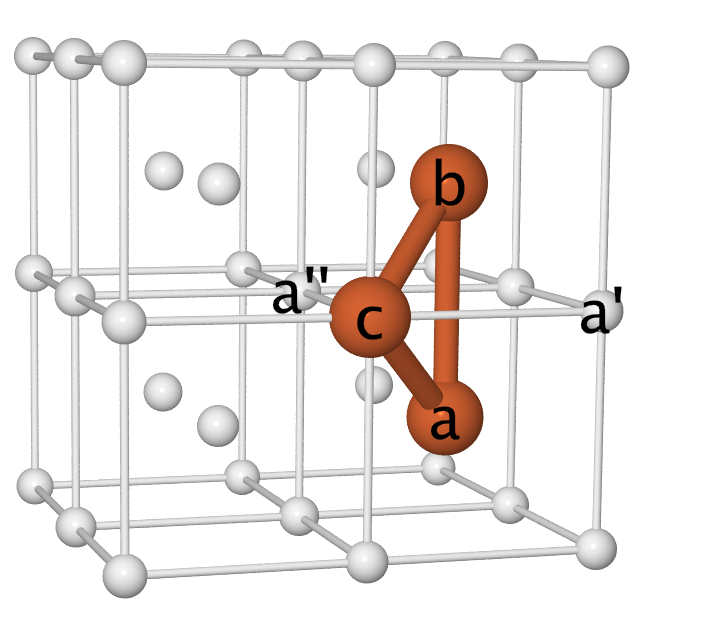}%
    \label{fig:3vacdiff_a}}%
  \hfill%
   \subfloat[Mobile configuration v$_3^m$]{%
    \includegraphics[width=4.3cm]{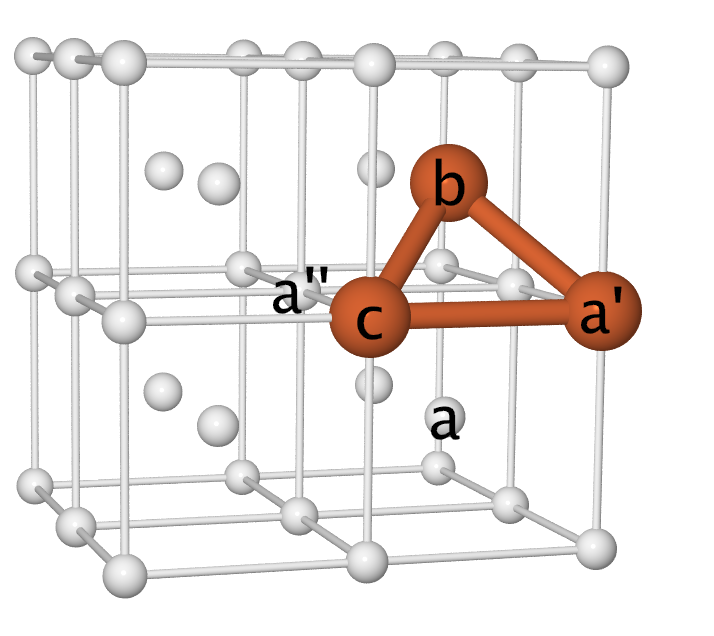}%
    \label{fig:3vacdiff_b}}\\
  \subfloat[Intermediate configuration]{%
    \includegraphics[width=4.3cm]{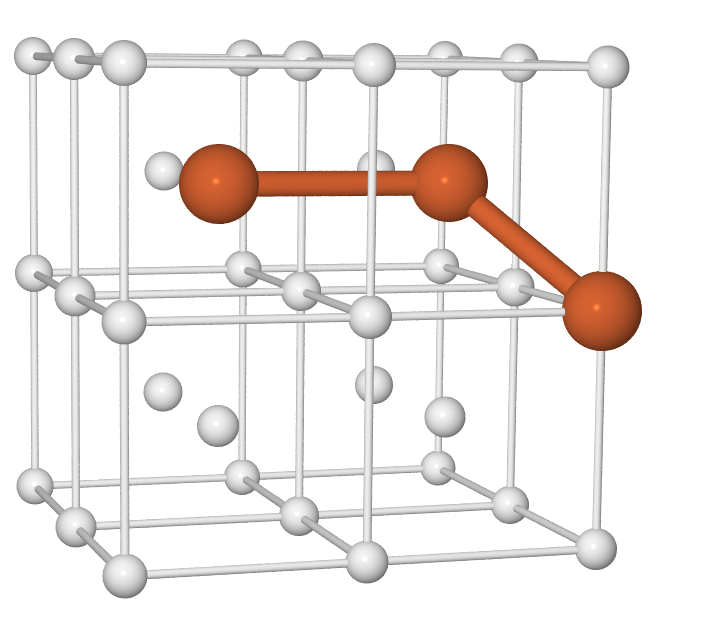}%
    \label{fig:3vacdiff_c}}%
\hfill%
  \subfloat[Immobile configuration v$_3^i$]{%
    \includegraphics[width=4.3cm]{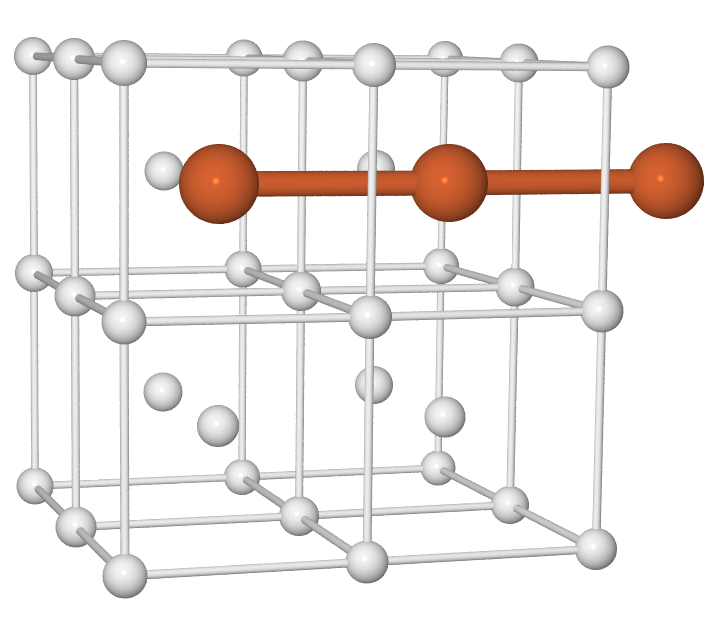}%
    \label{fig:3vacdiff_d}}%
  \caption{(Color online) \label{fig:3vacdiff}
 Three-vacancy cluster diffusion. 
 Small light spheres are occupied lattice sites, large dark spheres are vacancies.}
\end{figure}

\begin{figure}[hb]
  \includegraphics[width=\columnwidth]{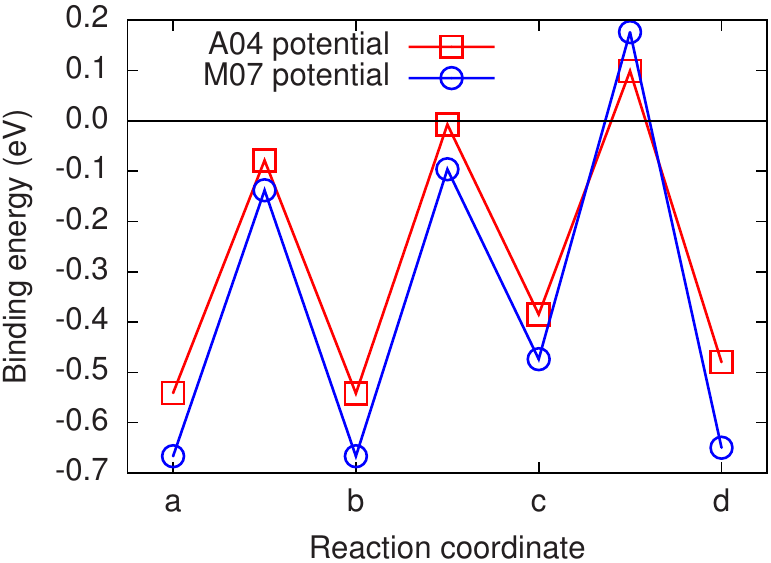}
  \caption{(Color online) Binding energy of minima and saddle points along the a--b--c--d trajectory (cf.\ Fig.~\ref{fig:3vacdiff}). Note that k-ART does not provide any energy information for intermediate points.\label{fig:3vac_traj} }
  
\end{figure}

In the mobile ground state configuration v$_3^m$ (bound by -0.54 eV [-0.67 eV] for A04 [M07]), vacancies $a$ and $b$ at 2NN separation have a third vacancy $c$ on a common 1NN site (see Fig.~\ref{fig:3vacdiff}\subref{fig:3vacdiff_a}).
The cluster diffuses by a 1NN vacancy hop of either $a$ or $b$ to one of the two sites that are 1NN to $b$ or $a$ and 2NN to $c$.
In Fig.~\ref{fig:3vacdiff}\subref{fig:3vacdiff_a} $a$ or $b$ could hop to either sites $a'$ or $a''$.
Figure \ref{fig:3vacdiff}\subref{fig:3vacdiff_b} shows one of the two possible results.
For the A04 (M07) potential the barrier for this event is 0.46 eV (0.52 eV).
This mobile v$_3$ configuration can transition to an immobile v$_3^i$ configuration, which is located 0.06 eV (0.02 eV) above the ground state, by a multistep process (cf.~Fig.~\ref{fig:3vacdiff}\subref{fig:3vacdiff_c}--\subref{fig:3vacdiff_d}).
However, the barriers to leave this metastable state are significantly
higher; at 573 K this results in a mean residence time that is longer
by a factor of about 7 (5) with respect to the mobile state. 
Figure~\ref{fig:3vac_traj} presents the energies and barrier heights along the trajectory shown in Fig.~\ref{fig:3vacdiff} for the two potentials.

For larger clusters, our simulations yield significantly longer lifetimes and lower diffusion coefficients than the lattice KMC simulations by Borodin \etal \cite{Borodin:2007:161}
This seems to imply that those simulations are not able to fully capture the complete dynamics of the system, leaving aside important diffusion pathways or neglecting elastic effects that are taken into account using k-ART.

\subsection{Six vacancy aggregation}
\label{sec:vacancy-aggregation}

\begin{figure}
   \includegraphics[width=\columnwidth]{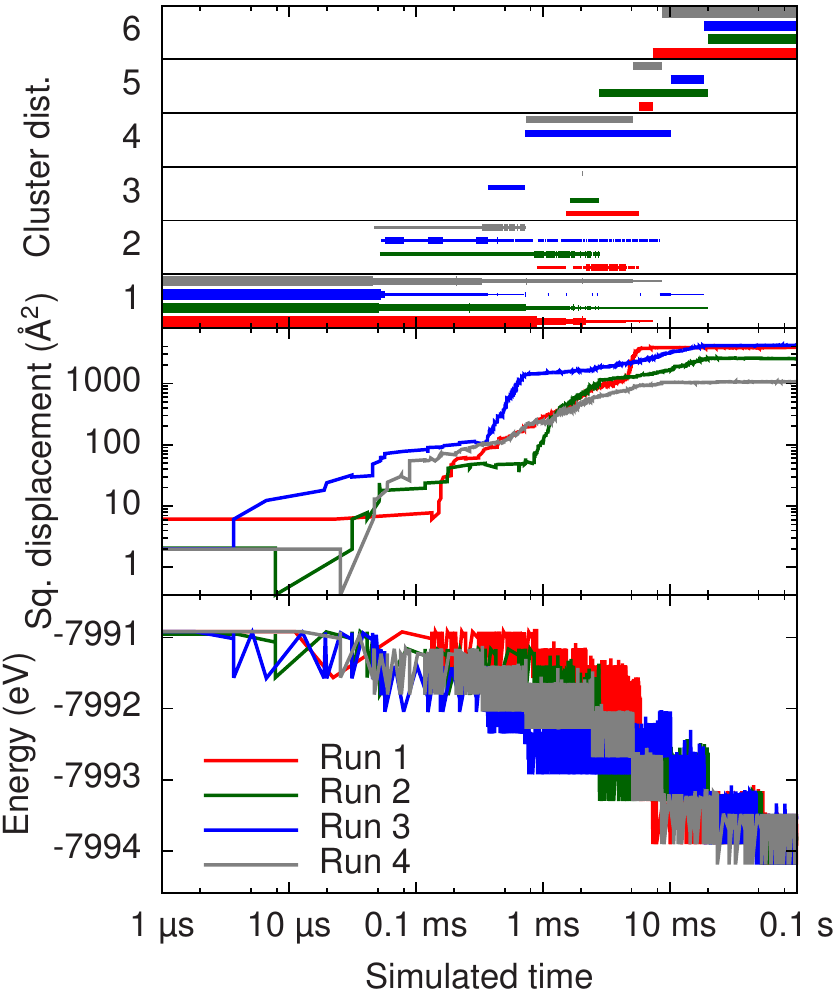}
   \caption{(Color online) Trajectory of four independent runs with 6 randomly placed vacancies.
     The bottom plot shows the cohesive energy of the system over time.
     The central plot shows the total squared displacement.
     The top plot shows the cluster size distribution, where the line width is proportional to the number of vacancies in clusters of the respective size.
     High atomic mobility appears to be correlated with clusters of size three.}
   \label{fig:traj_6v}
 \end{figure}

In this section and the following one, we assign a vacancy to a cluster, if it is at nearest or next-nearest atom site separation from at least one other vacancy of that cluster.
At the lower simulation temperature, the 4NN vacancy, which is technically still bound, is rarely sampled, and it does not affect the overall aggregation evaluation whether or not that site is included in a cluster definition.
We now look at vacancy aggregation by removing six atoms at random positions from the 2000-atom box.
The ground state for this system, at 2.63 eV below six isolated vacancies, is a single cluster, with four vacancies sitting at the corners of one face of the cubic unit cell, and the other two in the body centers of the two cubic unit cells that share this face.
Four independent simulations with the A04 potential are launched at 50 $^\circ$C, in the vacancy-dominated diffusion regime, until vacancies are completely agglomerated (1600--5400 KMC steps not counting the intrabasin flickering moves) for about 0.1 seconds.
The time evolution for these clusters is presented in Fig.~\ref{fig:traj_6v} and shows the evolution of the cluster size distribution, the total square displacement and the energy.
We note that even though some dimers are formed very early on, as soon as 50 \textmu s, diffusion becomes important only after almost 0.5 ms.
It is completed typically within 20 ms, when the full 6-vacancy cluster is formed.

 \begin{figure}
   \includegraphics[width=\columnwidth]{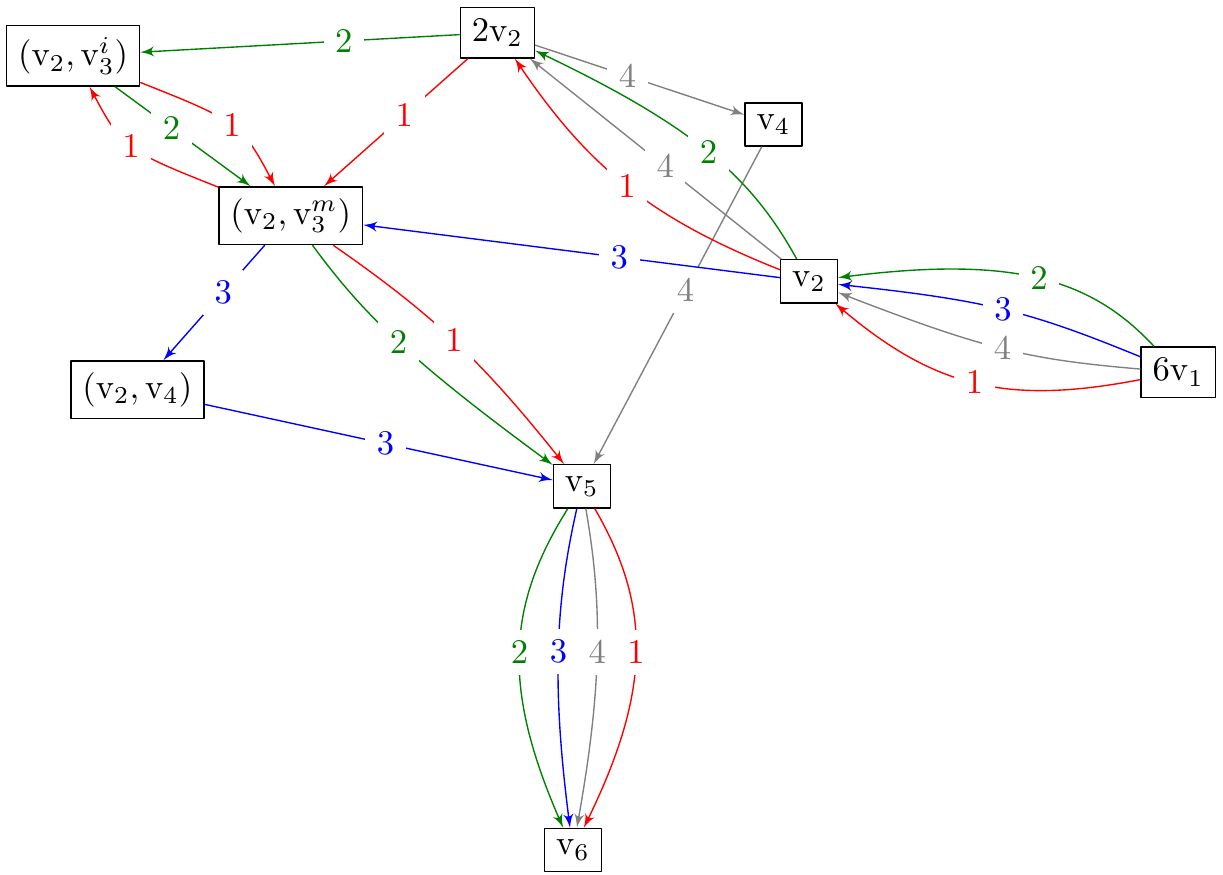}
   \caption{(Color online) Flow chart of vacancy cluster formation in the 6 vacancy system. 
     Clusters with $n$ vacancies are labeled v$_n$, monovacancies are not considered except for the initial state.}
   \label{fig:6v_flow}
\end{figure}

\begin{figure}
  \includegraphics[width=\columnwidth]{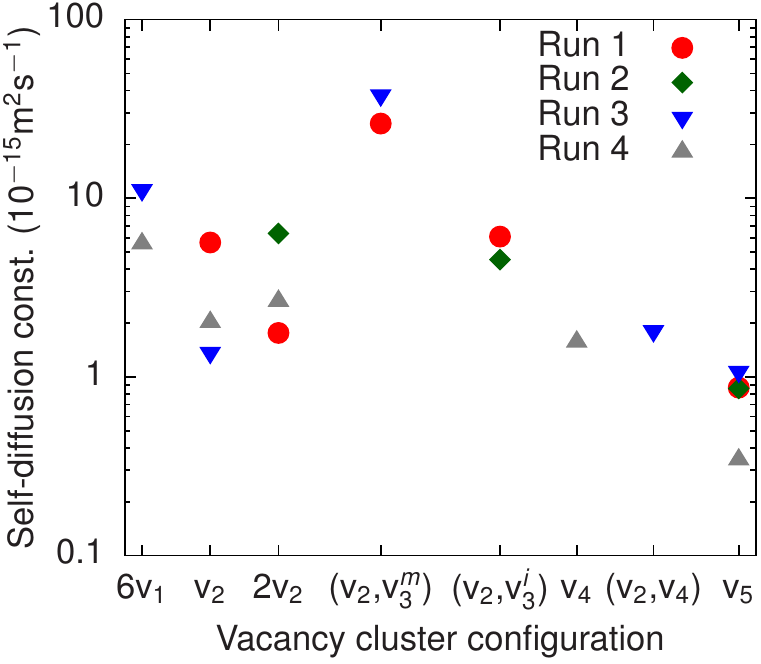}
  \caption{(Color online) \label{fig:difftime}
     Top: diffusion rate of various cluster combinations in six-vacancy
    simulations. Clusters of $n$ vacancies are denoted as v$_n$. Remaining vacancies are
    monovacancies. There are two distinct configurations of three-vacancy clusters:
    mobile (v$_3^m$) and immobile (v$_3^i$), see text. }

\end{figure}

As is seen in Fig.~\ref{fig:6v_flow}, all simulations show a similar aggregation pathway, with the formation of a first divacancy, followed by a second (skipped in run 3) and then the formation of more complex trivacancy, a divacancy with a tetravacancy, a pentavacancy and, finally, the ground state.
Overall, the diffusivity is about $10^6$ \AA$^2$/s but certain clusters stand out.
This is the case for the mobile trivacancy v$_3^m$, which shows the fastest diffusivity, while the pentavancy is extremely slow (see Fig.~\ref{fig:difftime}, which plots the iron self-diffusion constant for various cluster configurations).
This agrees with the findings from the 3-vacancy simulations presented in the previous section and can be observed in Fig.~\ref{fig:traj_6v}, where the fastest increase in total square displacement occurs in the presence of trivacancy. 
Run 4, which does not sample a trivacancy, shows a final square displacement that is between 3 and 4 times smaller than the three other runs, and confirms the rich pathway diversity to be found even for such a small cluster.

\subsection{50 vacancy aggregation}
\label{sec:50-vacancy-aggr}

\begin{figure}
  \includegraphics[width=\columnwidth]{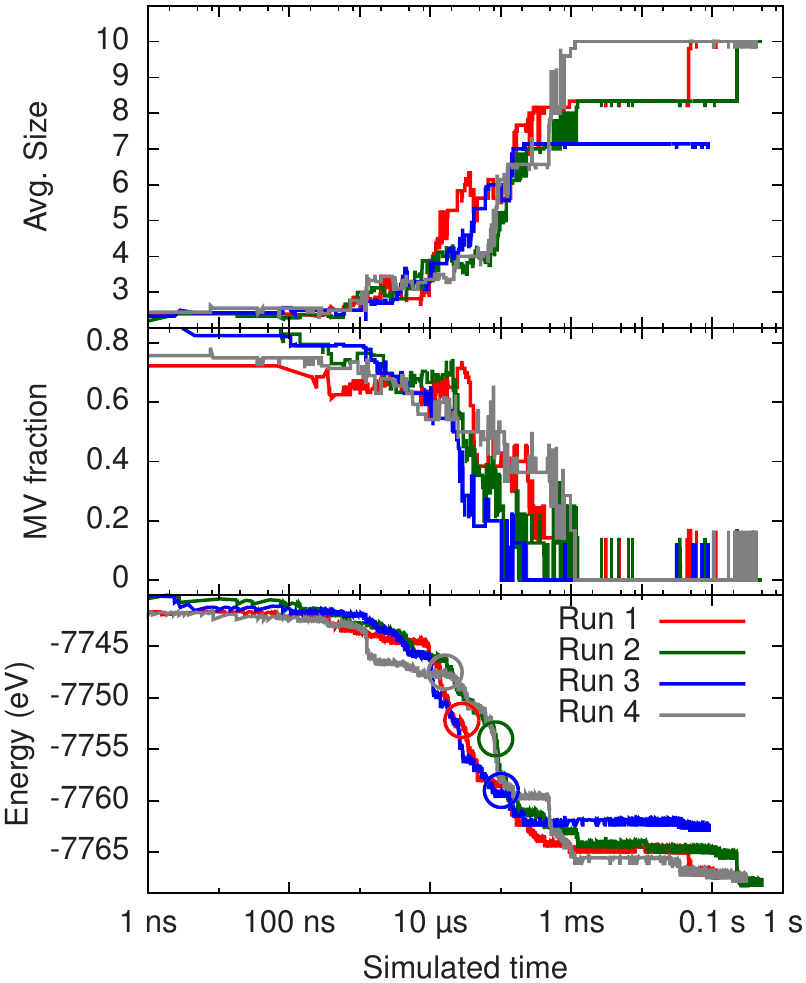}
  \caption{(Color online) 
    K-ART simulation results of four independent runs with randomly placed vacancies.
    Bottom plot: energy of the four runs. 
    Most energy is released within the first 100 \textmu s. Circles mark the highest effective barrier of the runs (see  Fig.~\ref{fig:50vmagn} for magnifications). 
    Center plot: fraction of monovacancies.
    After about 1 ms, all vacancies are clustered.
    Top plot: average cluster size.
    The simulations are terminated when seven (run 3) or five (other runs) clusters remain and no further progress is made.}
  \label{fig:traj_50v}
\end{figure}

\begin{figure}
  \includegraphics[width=\columnwidth]{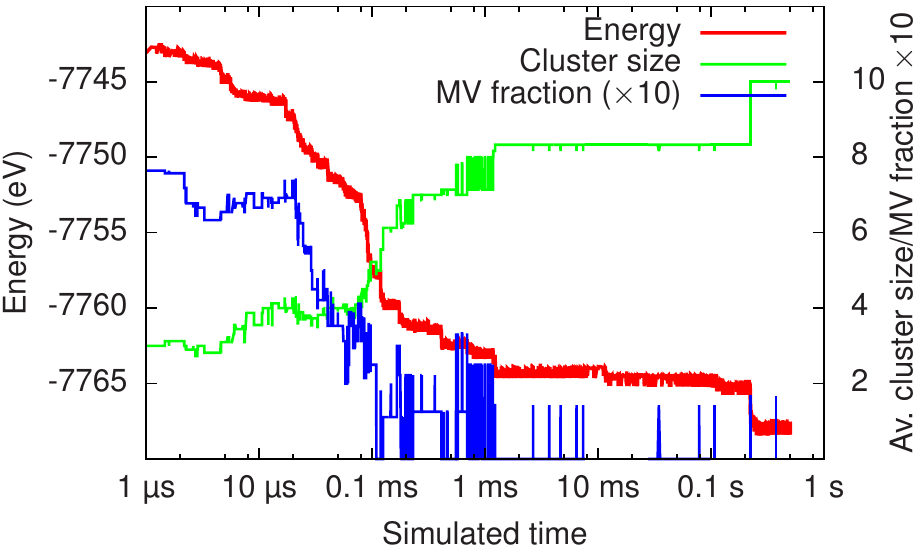}
  \caption{(Color online) Details of run 2: Energy, average cluster size and monovacancy fraction are plotted over simulation time. 
    For further details, see text.}

  \label{fig:traj_b}
\end{figure}

Having established the diffusion mechanisms for small aggregates, we can now see what role they play in an environment with a larger concentration of defects.
We prepare four independent models of 2000-atom Fe cell with 50 vacancies each inserted at random.
Fig.~\ref{fig:traj_50v} shows the trajectory for these four runs, with a similarly energy profile associated to three regimes: 0 to 10 \textmu s, 10 \textmu s to 1 ms and from 1 ms to the end of the simulation at 1 s.
To understand these three regimes, it is useful to look at the evolution of the average cluster size and the monovacancy fraction shown in Fig.~\ref{fig:traj_b} for Run 2.

\begin{figure}
  \includegraphics[width=.47\columnwidth]{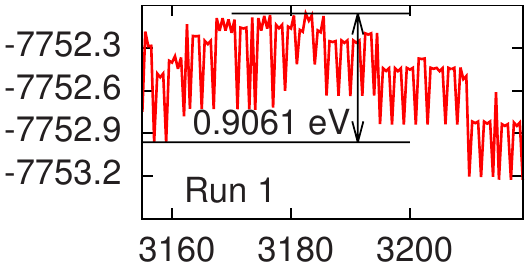}
  \includegraphics[width=.47\columnwidth]{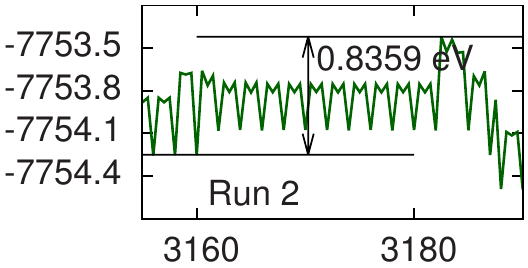}\\
  \includegraphics[width=.47\columnwidth]{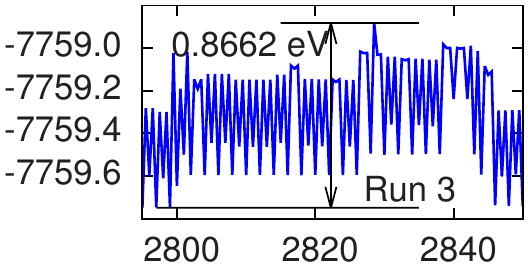}
  \includegraphics[width=.47\columnwidth]{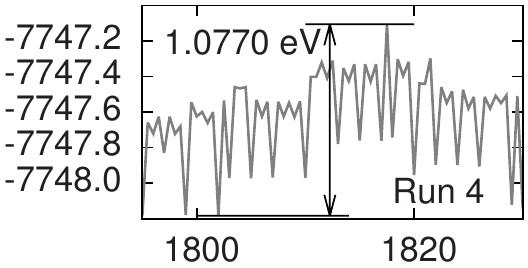}
  \caption{(Color online) Magnifications from the four runs showing the trajectory through minima and saddle points.
    The highest effective barrier (i.e.~difference between saddle point energy and previous lowest energy) is highlighted.
    In all four plots, the abscissas are KMC steps, while the ordinates are energy in eV.
    For run 1 and 2, an energetically favorable configuration is obtained within a few KMC steps.
    For runs 3 and 4, the energy does not drop below the former optimal energy right away.
    This is also an example of the ``Replenish and Relax'' mechanism described in Ref.~\onlinecite{Beland:2013:105502}.}
  \label{fig:50vmagn}
\end{figure}

The first regime is associated with internal cluster relaxation, with an almost constant cluster size and number of vacancies.
The second regime shows clustering, with the fraction of monovacancies dropping to almost zero (except for transient defects) followed by sharp increase in average cluster size.
This leads to a significant energy relaxation.
The final regime shows the slow internal evolution of larger clusters, that vary in size between 5 and 14, with an average of 8 vacancies, in agreement with positron annihilation spectroscopy measurements on iron samples.\cite{Eldrup:2003:346} 
After almost one second, the relaxation is still 20 eV above the lowest-energy structure, $-7789$ eV, found for a single 50-vacancy cluster relaxed using k-ART, suggesting that further aggregation takes place on a longer timescale at 300 K.

The time scale of vacancy relaxation is almost logarithmic.
For run 2, for example, 215 KMC steps are needed to reach 1 \textmu s, 650 KMC steps to reach 10 \textmu s, 3360 for 0.1 ms, 5000 for 1 ms, 6600 for 10 ms, and 7500 to reach 0.1 s (total 10600 KMC steps for a 0.5 s simulation, other simulations are comparable).
This is only possible because the effective barrier, i.e.~the one that leads to a structural evolution, increases roughly logarithmically with time.
The effective barrier is defined as the energy difference between the local saddle point and the lowest energy obtained previously (cf.\ Fig.~\ref{fig:50vmagn}, which shows the maximal effective barriers of the four runs in the relaxation phase).
For relaxation simulations, this is a good measure of the time scale, as this determines the time scale to escape from the previous optimal configuration.

\begin{figure}
  \includegraphics[width=\columnwidth]{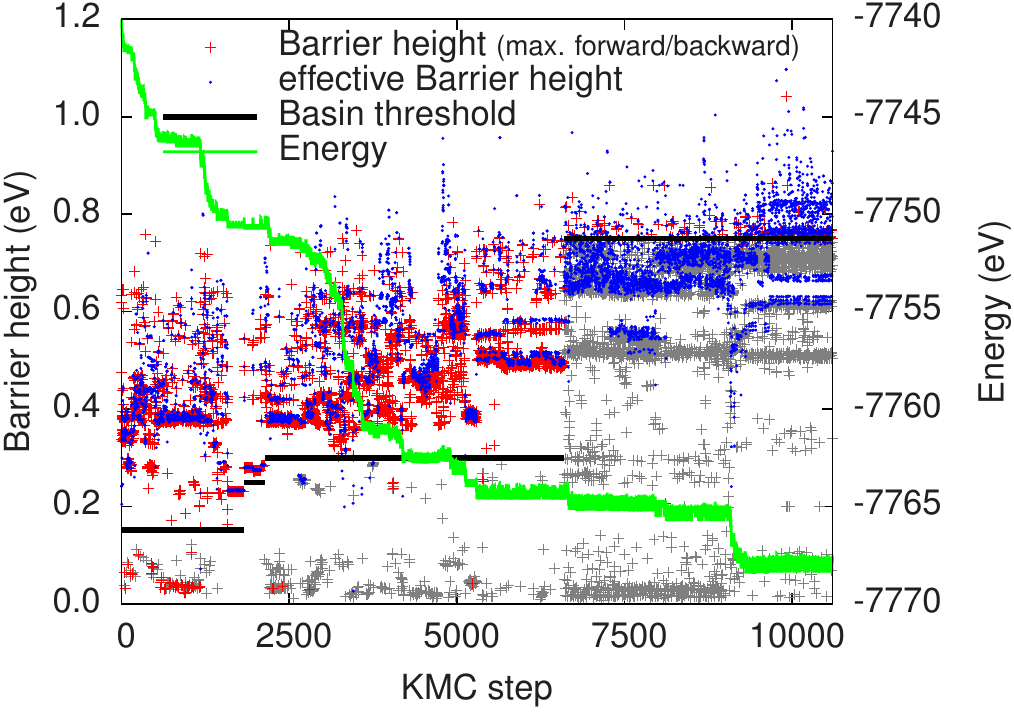}
  \caption{(Color online) Basin threshold control and relaxation progression of run 2.
    The crosses show the height of the saddle point above the lower of the two states that it links.
    Red symbols are regular barriers, gray symbols represent barriers included in the basin.
    The smaller blue symbols show the effective barrier, i.e.~the height of the saddle point above the minimal state obtained so far.
    The basin threshold is shown by the solid black line. }
  \label{fig:barr_basin}
\end{figure}

The efficiency of bac-MRM allows us to follow the overall relaxation over a long time scale.
An example, taken from run 2, is given in Fig.~\ref{fig:barr_basin}. 
Crosses indicate barriers selected: grey crosses are basin exploration events that will not directly advance a simulation; red crosses are events outside of basins and may lead to progress or return to previous state. 
Around step 2000, we observe a flicker associated with barrier of 0.23~eV (and no more basin events). 
We increase the basin threshold to 0.25 eV, just to find a flicker at 0.28 eV. 
We must therefore increase the threshold again to allow the simulation to progress. 
The new threshold, set at 0.3~eV remains effective until step 6600, when it is increased to 0.75 eV.

The blue dots in Fig.~\ref{fig:barr_basin} show the effective barrier height.
This energy dictates on which time scale a relaxation takes place, while the local barrier height is closely related to the time increment per KMC step.
This demonstrates the need for an effective basin acceleration scheme, as without bac-MRM the simulation would get locked in on the significantly shorter time scale associated with the lower local barriers associated with non-diffusive events.

During the course of this run, low effective barriers die out, except when there are large relaxations.
In these situations, we are finding new minimal energy states frequently, including skewed events, with small forward and large backward barriers.
Overall, the effective barrier increases roughly linearly with number of KMC steps or logarithmically with time.

The 50-vacancy simulations were stopped when the vacancies aggregated in 5 clusters (7 clusters for run 3), as no further progress could be seen.
The smallest cluster in this stage was size 6 (run 3: 5), whose effective migration barrier is significantly above the basin threshold of 0.75 eV at the end of the simulations.
To go further in time, beyond the 1-second time scale, it would therefore be necessary to further increase the basin threshold, to handle properly the high-barrier flickers that appear in this time regime.  
In the current bac-MRM implementation, all events for all atoms in all basin states need to be kept in memory, and the basin would comprise several hundreds, if not thousands of states for higher basin thresholds.

Aggregation of vacancies in this system has been used to compare the efficiency of two other off-lattice KMC methods looking, for example, at the time needed to reach an energy level of $-7760$~eV.
Using the autonomous basin climbing (ABC) method combined with nudged elastic band (NEB) and KMC methods, Fan \emph{et al.}\ obtain a relaxation time scale of more than 20~000 seconds,\cite{Fan:2010:104102} compared with 10$^{-4}$ to 6$\times 10^{-5}$~s with k-ART.
This indicates that ABC is unable to identify the most important relaxation pathways, significantly overestimating the relaxation time scale by missing relevant mechanisms.\cite{Brommer:2012:219601,Xu:2013:66} 
With the self-evolving atomistic kinetic Monte Carlo (SEAKMC) method\cite{Xu:2011:132103,Xu:2012:102} and running 12 simulations, Xu \emph{et al.} find a relaxation time between \num{3e-6} and \SI{e-4}{\second}.\cite{Xu:2013:66} 
The relaxation timescales overlaps with our results, but are within an interval about an order of magnitude faster than ours --  a small difference given that the timescale depends exponentially on the determination of barrier height.
Nevertheless, since Xu \emph{et al.}\ replicate carefully our simulation set-up, it is important to identify the origin of this small discrepancy and efforts are currently underway to address this issue.
From our current and joint assessment with Xu and Stoller, both SEAKMC and k-ART provide the correct kinetics and the difference is due to small details in the implementation and choice of parameters that do not affect the overall results.

\subsection{Long-range elastic effects}
\label{sec:ge-results}
\begin{figure}
  \includegraphics[width=\columnwidth]{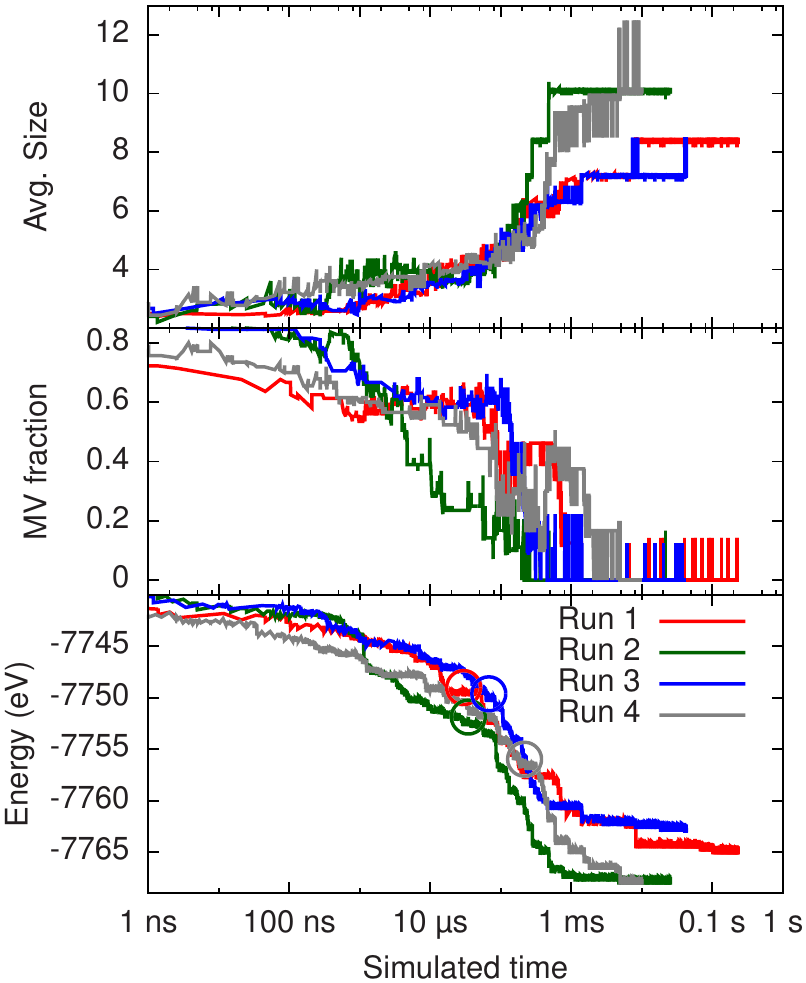}
  \caption{(Color online) 
    GE k-ART simulation results of four independent runs with randomly placed vacancies.
    Bottom plot: energy. 
    Circles highlight the highest effective barrier of each run (0.858/0.897/0.823/0.903 eV).
    Center plot: fraction of monovacancies.
    Top plot: average cluster size.
}
  \label{fig:traj_50_ge}
\end{figure}

We repeated the four runs with 50 vacancies using the generic embedding k-ART code, thus purposefully ignoring any elastic effects on defects beyond the topological classification radius of \SI{5.6}{\angstrom}.
The runs were stopped if no further progress was made.
The resulting trajectories are shown in Fig.~\ref{fig:traj_50_ge}.
Overall the differences to a full k-ART run appear to be minor: Two runs finish up with 6 clusters and a final energy above \SI{-7765}{\electronvolt}, and two with 5 clusters and below \SI{-7765}{\electronvolt} (run 4 briefly samples 4 clusters, with two close clusters temporarily connected via a second-nearest neighbour site, before settling with two separate clusters -- this might be an effect of GE). 
This is in agreement with simulations performed with SEAKMC\cite{Xu:2013:66}, a method that involves cutting off elastic interactions at roughly 8~\AA in the case of Fe vacancies, which exhibit similar trajectories. 

There are real differences, however, mostly in the intermediate time regime. First, due to the absence of long-range elastic contributions monovacancies appear to remain present until after the first ms, whereas they do no longer factor in full k-ART simulations at that time. 
Final structures show comparable distribution of cluster sizes in both cases.

\begin{figure}
  \includegraphics[width=\columnwidth]{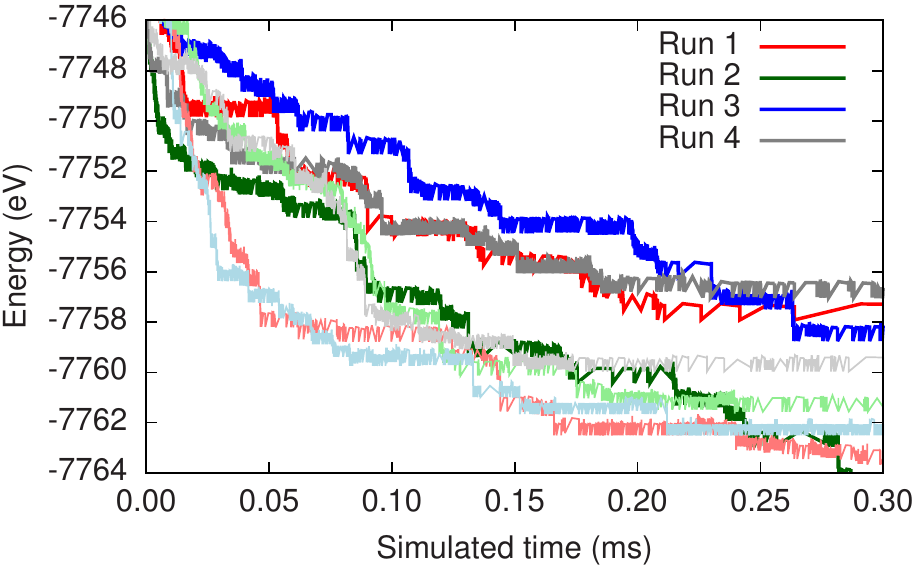}
  \caption{(Color online) 
    Comparison between standard and generic embedding k-ART runs from same initial configurations.
    GE runs are shown in dark colors and bold lines, whereas the corresponding standard runs use thinner lines and lighter colors.
    The time axis is non-logarithmic.
}
  \label{fig:eng_50_ge}
\end{figure}

Second, we observe differences in the  trajectories in Fig.~\ref{fig:eng_50_ge}. 
During the clustering regime (cf.\ Sec.~\ref{sec:50-vacancy-aggr}) after the initial relaxation, all full k-ART simulations show a pronounced drop in energy of about \SI{4}{\electronvolt} within a few microseconds, where the energy release rate accelerates after first slowing down.
In contrast, this behaviour is found only in run 2 of the GE simulations.
The other three runs yield an almost linear time-energy relationship during this regime of vacancy clustering, with a continuous and gradual release of energy.
This effect could be explained by the inherent symmetry of GE simulation runs: 
Without elastic effects, the quick cascades seen in full simulations are no longer favored, and the energy decreases more uniformly.

This does not seem to lead to significant effects on the long-term evolution, where these differences on the \si{\micro\second}--\si{\milli\second} scale become irrelevant, as the final vacancy aggregation limits long-range elastic effects and the GE approach still treats exactly the short range (\SI{5.6}{\angstrom}) elastic deformations.
As we compute the energy correctly in both methods, the final states reached are more determined by thermodynamics than by kinetics.
It should also be remarked that even with generic embedding the simulations are still considerably more sophisticated than with a standard, short-range, on-lattice KMC method, as even the artificially simplified GE simulations still allow complicated and off-lattice atomic defect structures.

\section{Discussion and conclusion}
\label{sec:conclusion}

Using the kinetic Activation-Relaxation Technique, an off-lattice kinetic Monte Carlo method with on-the-fly catalog generation, we study vacancy diffusion and aggregation in bcc Fe providing detailed information regarding mechanisms and pathways that is difficult to obtain through standard simulation methods. 
Our results show a richness in the diffusion mechanisms as well as a complex balance between elastic and chemical effects.

Comparison with previously published works of the mono- and divacancy diffusion, using two different forcefields, allow to show that k-ART samples correctly diffusion mechanisms, including artificial ones induces by one of the forcefields.\cite{Malerba:2010:19,Djurabekova:2010:2585}

First considering small vacancy cluster diffusion, with one to six defects, we can compare with MD simulations as well as provide a clear description of diffusion and assembly mechanisms.
For example, we observe that trivacancies diffuse an order of magnitude faster than mono-, di- and tetravacancies, as had been predicted by \emph{ab-initio} calculations of dominant cluster diffusion mechanisms for one to four vacancies.\cite{Fu:2005:68} 
Here, however, we go beyond these results and show for example, that penta-vacancy clusters are between 2 and 10 times slower and larger clusters, counting 6 to 14 vacancies, are almost immobile on the simulation time scale.
The identification of pathway and diffusion mechanisms for these structures are therefore essential for understanding the long time kinetic of materials.
Methods such as k-ART, which includes both elastic and kinetic effects, greatly facilitate this characterization as compared to standard approaches where these pathways must be deduced one by one, an approach that rapidly becomes impossible in when dealing with long-range elastic effects or complex systems.

Simulations of the kinetics of 50 vacancies in Fe were performed at \SI{50}{\celsius}, in the vacancy-migration regime discussed by Fu \emph{et al.}\cite{Fu:2005:68} Here, we observe the aggregation in on average 10-vacancy clusters, on a timescale between \SI{1}{\milli\second} and \SI{1}{\second}, depending on the specific initial vacancy distribution.
Once these clusters are formed, we do not observe further aggregation on the simulation time scale.
This observation is in agreement with position-annihilation spectroscopy experiments that show that room-temperature annealed Fe-implanted samples still present a signal compatible with clusters of 9 to 14 vacancies in size at room temperature.
On the other hand, larger voids with 40 to 50 vacancies are observed only after annealing at temperatures between 100 and \SI{280}{\celsius}.\cite{Eldrup:2003:346}

Comparative simulation runs, where long-range elastic interactions were artificially excluded, show that elastic effects are particularly significant during the clustering phase up to about 0.3 ms. 
There, these effects trigger quick energy release cascades that are not regularly seen otherwise.
On longer time scales, long range elastic interactions do not appear to influence either defect evolution or final states reached.
As the elastic strain associated with vacancies in a compact material like bcc iron decays rapidly with distance, it is to be expected, that elastic effects play an even larger role for interstitials or in more open (e.g.\ covalently bound) systems.
This shows the importance of including these interactions in the determination of energy barriers and thus simulation kinetics.

Beyond providing detailed information regarding the long-time kinetics of vacancies in Fe, this work opens up a new time regime to simulations, expanding the overlap between simulations and experiments and technological applications. 
Kinetic ART, which can treat off-lattice positions and disordered materials while including exactly long-range elastic effects, will allow us to further our study of defects kinetics in Fe, but also in a number of other technologically relevant materials, such as alloys, semiconductors and cements.
In these materials, mechanical and electronic properties are determined by microscopic kinetics and structure, and we can now answer those questions, which have remained out of reach until today.

 \begin{acknowledgments}

   This work has been supported by the Canada Research Chairs program and by grants from the Natural Sciences and Engineering Research Council of Canada (NSERC) and the \textit{Fonds de recherche du Québec -- Nature et technologies} (FRQ-NT).
   We are grateful to \textit{Calcul Québec} (CQ) for generous allocations of computer resources.

The kinetic ART software is available upon request to the authors.
	
 \end{acknowledgments}
This is a post-print of Ref.\ \onlinecite{Brommer:2014:134109}. \copyright\ 2014 American Physical Society.

\bibliography{kART}

\end{document}